# NEURAL NETWORKS AND DATABASE SYSTEMS

Erich Schikuta

University of Vienna

Department of Knowledge and Business Engineering

Rathausstraße 19/9, A-1010, Vienna, Austria

*Abstract*

*Object-oriented database systems have proven very valuable at handling and administrating complex objects. In the following guidelines for embedding neural networks into such systems are presented. It is our goal to treat networks as normal data in the database system. From the logical point of view, a neural network is a complex data value and can be stored as a normal data object.*

*It is generally accepted that rule-based reasoning will play an important role in future database applications. The knowledge base consists of facts and rules, which are both stored and handled by the underlying database system.*

*Neural networks can be seen as representation of intensional knowledge of intelligent database systems [14]. So they are part of a rule based knowledge pool and can be used like conventional rules. The user has a unified view about his knowledge base regardless of the origin of the unique rules.*

## 1   Introduction

In this paper a framework for the modelling of neural networks and their integration into database systems is presented. This framework is based on the object-oriented approach, which is provided by many modern database systems. It allows for the easy and handy definition and administration of neural networks of predefined database types or the creation of new and specialized network types. To model neural network objects the generator approach of Smith [22] is used, which is extended by communication operators for the database environment. Different operator models are described and their properties discussed. Further we present a common framework for embedding neural networks into the rule based component of a knowledge-based database system.

It is our goal to treat the nets like normal data in the database system. Hereby we follow the vision of G. Vinek [24]:

"*With the development of (oo) applications ... it has to be ensured that instantiated objects have to resemble real objects ... of the application area in structure and behaviour.*" (text translated by the author)

According to this, a neural net is a complex data value and can be stored as a normal object. The object-oriented approach seams (and in our opinion has proven) the most comfortable and natural design model for neural networks [10]. In the terms of object oriented database systems neural networks are treated generally as complex objects. These systems showed very valuable at handling and administrating such objects in different areas, as computer aided design, geographic databases, administration of component structures, and so on.

The proposed paradigm allows the definition of a single network up to a family of similar networks. These are networks with similar properties, like network paradigm and structure. The term similar is not restricted to network properties only, but covers also networks, which accomplish special tasks. The characteristics of these networks can be very different, but the database input (the task description) is the same. The data manipulation facilities of the system are exploited to handle the networks and their tasks within the same framework.

The usage of a database system as an environment for neural networks provides both quantitative and qualitative advantages.

*Quantitative Advantages*: Modern database systems allow an efficient administration of objects. This is provided by a smart internal level of the system, which exploits well-studied and known data structures, access paths, and more. A whole bunch of further concepts is inherent in these systems, like models for transaction handling, recovery, multi-user capability, concurrent access, to name only a few. This gives an unchallenged platform in speed and security for the definition and manipulation of large data sets.

*Qualitative Advantages*: The user has powerful tools and models at hand, like data definition and manipulation languages, report generators or transaction processing. These tools provide a unified framework for handling neural networks and the input or output data streams of these nets. The user is confronted with one general user interface only. This spares him awkward tricks to analyze the data of his database with a separate network simulator system.

A second very important aspect is the usage of neural networks as part of the rule component of a knowledge-base database system [15]. Neural networks represent inherently knowledge by the processing in the nodes [13]. Trained neural networks are similar to rules in the conventional symbolic sense. A very promising approach is therefore the embedment of neural networks directly into the generalized knowledge framework of a knowledge-based database system.

## 2  The Neural Network - Database System Integration Framework

One valuable property of neural nets is their beneath unrestricted flexibility in providing solution strategies to all types of problems. This leads from different versions of neural nets of one net paradigm for different problems to nets of different paradigms for a common problem. The data model of the underlying object-oriented system has to be based on three elements: types, object, and functions.

Objects represent unique entities and are classified by types and described by functions (correspond to attributes in the relational model). Types are organized in type hierarchies with function inheritance. Functions access and manipulate objects and can be applied to other functions. This indirection allows a high degree of flexibility and expressional power.

## 2.1 The neural network types and objects

One describing property of the object-oriented design is the hierarchy of types. A type comprises a set of objects, which share common functions. Generalization and specialization define a hierarchical type structure, which organizes the unique types. All of its subtypes along the type hierarchy also inherit functions defined on a supertype.

A paradigm for neural networks in database systems has to provide flexibility in two directions, embedment level and expressional power.

*Embedment level*: The system has to support the user on all levels of embedment [16]. The term embedment was introduced to define the user control of the static and dynamic properties of neural network objects. Control means the possibility to change or adapt the properties in question. Existing neural network systems can be classified according to the embedment level.

*Expressional power*: The proposed framework has to cover all different types of networks. This reaches from simple networks consisting of a few processing elements to large net systems, which comprise several networks (a network of networks) possibly with different net paradigms. In the last few years these systems yield specific importance, e.g. in the area of data analysis, image processing, etc., where different networks are closely connected and responsible for different tasks, like noise reduction, pattern recognition.

The object-oriented system has to provide a generic type for a neural network system. This basic generic neural network type comprises all possible network paradigms and embedment level. We call this type NUnit (for Neural Unit). The purpose for this rather generic type is the possibility to express complex neural systems consisting of different networks. Further, it allows for the definition of the data communication with the environment, in our case the database system. This results in data input and output objects. We use the term object to avoid the anticipation with a specialized, implementation dependent, data format, like functions and streams.

The NUnit type is a subtype of the general object type. Subtype of this NUnit type is a generic neural network type, NeuNet. It represents a unique network, which defines the net paradigm, connected processing elements. Specializations of this type are predefined system types or customized, user defined types.

Figure 1 depicts the hierarchical structure of the neural network types.

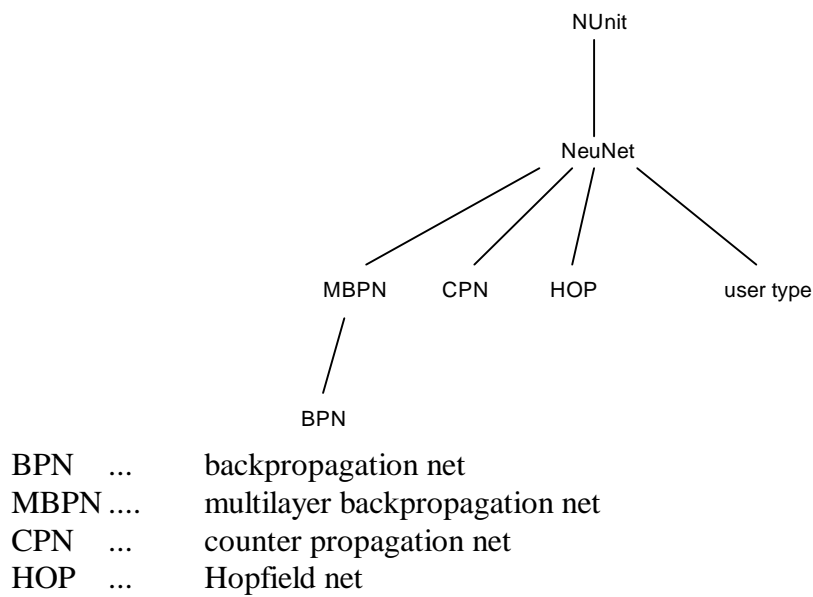

BPN ... backpropagation net
MBPN .... multilayer backpropagation net
CPN ... counter propagation net
HOP ... Hopfield net

Figure 1: type hierarchy of NUnit

The general NUnit type provides just the functions for name. A finalized subtype of NUnit, which defines the net paradigm, defines all other describing and necessary characteristics. At a very early moment the user can create an instantiation of a neural network system, but has not to decide about the number of networks, net paradigm, attributes, and functions. This is very useful for database systems where families of networks have to be administrated for similar problems. These neural network objects can be referenced by names as all other objects in the database. All data administration facilities of the system are applicable on NUnit objects as well, as insertion, deletion, and update.

Further the user has the possibility to define special properties for an instantiation of these data types, like processing elements, connections, local memory, transfer functions, and so on [9].

## 2.2 Formalism

In the following discussion we use the general terminology introduced by [22]. A neural network system consists of 2 element sets, nodes N and arcs A. The nodes describe the characteristic properties, like node names, ports, state-space and functions. A node n can be seen as a tuple

$$n \in \text{NODENAME} \times \text{set(PORTS)} \times \text{STATE-SPACE} \times \text{FUNCTIONS}$$

The function component defines all net functions algebraically, like activation, input or output function and all net dynamics, like evaluation and training method.

The arcs represent the connections between processing elements. These connections are directed and can be described by an outport and an inport. An arc is a tuple

$$a \in \text{OUTPORT} \times \text{INPORT}$$

Formally, a net is therefore described by

$$\text{Net} = (N, A)$$

where N is the set of nodes and A is the set of arcs

This description defines a whole family of similar networks, similar in respect to their describing characteristics. A generator function instantiates networks according to the description.

In the context of a database system this definition is not comprehensive enough. The environment of the network is an essential part of the whole framework. Smith is aware about the importance, but shifts the environment interaction to a specialized port set. This port set is not flexible enough for our framework and therefore we extend his formalism by an additional function set for the environment. This set contains function definitions for the input from and the output to the database system, the input resp. output operators. This operators can be described by a tuple

$$E \in \text{INPUT-OPERATOR} \times \text{OUTPUT-OPERATOR}$$

Accordingly, a neural system in our framework is defined by

$$Net = (N, A, E)$$

On each level of the proposed database type hierarchy, attributes are or can be defined for these describing sets. The super type NUnit contains functions for name and for the environment communication. The operators for the database input and output have to be defined at this level.

## 2.3 Type hierarchy

The type hierarchy expresses hereby its strength. The user has not to specify anything, because the choice of a provided system type predefines all these functions and operators. This means, if he chooses the predefined BPN type (see Figure 1), he can use and handle it with the accompanied functions and tool set of the type.

A further expressional advantage is the usage of a predefined type and the overloading of its operators. This comes very handy with the object-oriented approach. Functions or operators can be defined for an existing subtype and replace the original ones of the super type. This allows the definition of highly complex neural systems and networks. For example, a system can be specified which has different dynamics in different layers.

As mentioned above a network can also be built from scratch. The hierarchical type system provides the model for the formal definition of a network family. For the realization of a specific network additional classes are necessary, classes for processing elements, connections, and layers. The user can define these freely at his/her will. Object identifiers provided by the database system identify the objects of a unique network. Via these identifiers the functions for the network dynamics can be defined.

## 2.4 NUnit operators

All networks obtain specialized functions to support the evaluation and training phase (these functions are inherited by the super type NUnit). We partition these functions into 2 groups,

- External net operators, and
- Internal net operators

In the following we use operators instead of functions to express the active component more accurately. Functions would refer to strongly to a functional behaviour, which would restrict the possible realizations. Also we try to separate the NUnit operators from the database functions.

### 2.4.1 External net operators

The external net operators build the communication to the embedding environment, in our case the database system. They are used to administrate the dynamic components of the network, like training phase and evaluation phase. Here we distinguish two operators:

- Input operator, and
- Output operator

The Input operator provides all incoming data. This can be the data set for the training phase or the input for the network evaluation. In the database context the data stream is defined by database query statements, which describe the accessed data within the database framework.

The Output operator handles all data, which is 'produced' by the neural network. This can be weight information after a successful training phase or the result of an evaluation phase. We see three different approaches to describe this Output operator:

- Automatic triggered insertion,
- Definition of an insert statement, and
- Return data stream of a database function

The automatic triggered insertion is a succeeding update of a produced neural network result triggered by a record insertion. A specialized evaluation class provides this functionality. It is built from one attribute for the insertion data stream and one for the output data stream.

Another possibility is the definition of a particular insert statement. It describes the insertion of the output data stream into a database class (or more classes).

The last possibility is similar to the return value of a mathematical function. The evaluation of specific neural network function produces an output data stream. The basic database tools, like the data manipulation language or the report generator, can direct this data stream.

These few special operators allow an easy usage of the neural network in conformance with the embedding database framework.

The training phase is simply an input operator. It resembling a special case, because this input triggers the dynamic training run of the neural network simulator. It can be simply modelled by an automatic triggered insertion. Another possible approach is by a special training operator, which extends the database query facility. However this necessitates an adaptation of the system for neural networks; in contrast to the automatic triggered insertion operator, which is actually constructed by the underlying database query system.

### 2.4.2 Internal net operators

The internal net operators define the dynamic behaviour of the neural net, like the network paradigm or the propagation algorithm. Generally, two approaches can be seen to define internal net operators,

- Dynamic definition, and
- Static definition

*Dynamic definition*

To allow the user to define the neural net paradigms dynamically, an appropriate functionality has to be provided by the database system framework. This can be achieved by definable operators, which are available in a number of existing database systems, like Iris [6] or Postgres [23].

To define dynamic methods on objects two different approaches are possible, via

- Foreign functions or
- Query language functions.

Foreign functions are functions, which are written in some general-purpose programming language outside the system. In this case the database compiler can not perform specific optimizations. Therefore, the programmer of the foreign function has to guarantee the correctness and optimization. Foreign functions have (hopefully) the advantage of faster execution speed. This is useful for extensive numerical calculations.

Query language functions are stored as data and are evaluated by a general-purpose neural net machine interpreting the paradigm data. A disadvantage is that the evaluation machine cannot exploit advantages of the underlying hardware system (like parallel processors, coprocessors, etc.).

*Static definition*

The different network paradigms are implemented directly as part of the database system. This gives the possibility to exploit special system properties. A trade-off of this approach is that neural net paradigms are supplied by the system and not generally definable by the user via the data manipulation language of the database.

## 3   Integration of Neural Networks into the Iris Database System

For justification of the described framework we present a practical integration of neural networks into the Iris database system [6]. The data model of the Iris database system is based on three elements: objects, types, and functions. Functions are expressed syntactically by the function keyword in type statements. All objects in Iris are handled with the global data manipulation tool of the system, the Object SQL interface (QSQL). OSQL is an object-oriented extension of the well-known SQL.

### 3.1   Structural definition of a Neural Net

A neural network in our model can be represented by the 3 basic object types we mentioned in the preceding section, NUnit, NEUNET, PElement. These types would be predefined in the system, but we want to give the OSQL definition for a comprehensive explanation.

```
Create type NUnit (Name  CharacterString);
```

This is the basic generic neural net object type. The only function is the name of the object, but it is arbitrarily extendible by the user. Every other net object is a subtype of NUnit.

The object type NEUNET represents the basic functional neural net unit type. It can be a whole network, a subnet, a layer or a group of similar processing elements only.

```
Create type NEUNET subtype NUnit
(
   InputData      Real many,
```

```
    CheckData       Real many,
    InputF          function,
    OutputF         function,
    ActivationF     function,
    LearnRate       Real;
    NeuronalUnit    NUnit many,
    UpdateOrder     <NUnit, Integer> many
    LearnOrder      <NUnit, Integer> many
);
```

The functions InputData and CheckData represent data streams. The neural network processes these during input in the evaluation or training phase or checking for target values during the training phase. The InputF, OutputF and ActivationF represent the transfer functions respective to the commonly known processing element model [4]. They reference the default evaluation functions of the processing elements of the network.

The multivalued function NeuronalUnit contains the neural net objects of the represented network. The UpdateOrder and LearnOrder functions result in binary tuples, which give the order of the neural net objects for the dynamic methods. This is based on the set property of an object oriented database system, where each set of elements has an arbitrary order. It is not suitable for the algorithms of some neural networks (e.g. Kohonen net, time function, backpropagation net, update order of the layers during the training phase, etc.). If no order function is defined, a system dependent arbitrary order is assumed.

The PElement object type defines the basic processing unit of a neural net. This object is a "terminal" object in the logical net hierarchy; that means, it does not consist of a group of neural net objects.

```
Create type PElement subtype of NUnit
(
   Activation      Real,
   InputF          function,
   OutputF         function,
   ActivationF     function,
   Predecessor     Link many
);
```

The PElement object contains the local memory function Activation, which stores the activation value of the unit. It is possible to define further storage functions (for more complex networks) or specialized transfer functions. Connections of a processing unit X is defined via the multi-valued Predecessor(X) function. It evaluates to a set of Link objects, which contain 3 functions giving the object identifier Y of the preceding neural object, the weight value and the delta value (update value during training phase) of the link. The object X is connected to all sub objects of Y.

```
Create type Link
(
   LinkFrom        NUnit;
   LinkWeight      Real;
   LinkDelta       Real;
)
```

## 3.2  Input and Output

The Input is done at the level of a NEUNET only. It can be interpreted as a simple stream of data. Therefore the InputStream-function of a NEUNET object can be a SQL-query or the output of another NEUNET object. In general nothing is said which output line is connected with which input line. Basically the order is system dependent and arbitrary. If a specific assignment of the input value order to the processing elements of a NEUNET object is necessary, the possibility of a PElement subtype IElement res. OElement (for input resp. output units) is given, where a specialized order of the units is definable.

```
Create IElement subtype of PElement (Order  Integer);
Create OElement subtype of PEleemnt (Order  Integer);
```

It has to be guaranteed that the number of input lines is equal to the number of output lines.

With these predefined definitions the system allows to define any possible net work structure. Because of the arbitrary extensibility of the object types, a suitable structure for any known neural network paradigm is producible.

In the following, we want to show how a neural network can be realized by a OSQL definition. Along with the descriptions of the various elements, we give a complete and comprehensive definition of a solution to the XOR-problem again using a feedforward multilayer network and the backpropagation training algorithm.

We use no predefined function. We define any necessary method from scratch using the OSQL database language.

First we have to define object instantiations for our network, which consists of 4 processing elements (Input1, Input2, Hidden, Output) arranged in 3 layers (Input, (the 2 one-unit layer) Hidden, and Output).

```
Create NEUNET (Name) instance XOR-Net("XOR-example");
Create IElement (order) instance Input1(1);
Create IElement (order) instance Input2(2);
Create PElement () instance Hidden;
Create OElement () instance Output;

Create NEUNET (name) instance Input("Input-Layer");
Set NeuronalUnit(XOR-Net) = (Input, Hidden, Output);
Set NeuronalUnit(Input) = (Input1, Input2);
```

The next step is to establish the network connections.

```
Set Predecessor(Input) = (Hidden, Output);
Set Predecessor(Hidden) = Output;
```

Finally, the general weight initialization values have to be defined.

```
Set Weight(XOR-Net) = 0.0;
```

**3.3   Definition of neural networks dynamics**

To define dynamic methods on objects two different approaches are possible, via foreign functions or query language (OSQL) functions as described above. For the definition of the update and training methods here, we use the OSQL formalism.

3.3.1   Update method

The update method defines the algorithm to calculate the neural net output to a given input. In our XOR-example we establish the commonly known feedforward execution, which can be described by the following equation

$$a_i(l) = f(\Sigma\, w_{ij}(l - 1, l) \cdot a(l - 1)), \quad l = 1, \ldots, L \text{ and } i = 1, \ldots, n_l$$

L is the number of network layers (3 in our case) and $n_l$ the number of processing units in layer l. $w_{ij}(l-1,l)$ denotes the weight of the connection between unit i on layer l and unit j on layer l+1. The activation value associated with unit i on layer l is $a_i(l)$. f is a nonlinear sigmoid function of the form $f(x) = (1+e^{-x})^{-1}$.

Different algorithms are definable for each neural net object. At least a general method has to be defined at the highest level, which is by default valid for all lower levels.

This property is exploited in our XOR-example, where two different versions of the InputF function of the feedforward algorithm have to be installed, one for the Input layer unit and one for the remaining processing elements. This is motivated by the different origins of the input values.

```
Set InputF(XOR-Net) = WeightSum;
Set InputF(Input) = WeightSumI;
Set OutputF(XOR-Net) = Ident;
Set ActivationF(XOR-Net) = Sigmoid
```

The values of the above functions are the names (or references) of other functions, which can be foreign functions or, like in our case, have to be supplied by the user.

```
Create function WeightSum(U NUnit) as
   begin
      Set Activation(U) = sum(listprod(
         select Activation(A) for each NUnit V where V in (
            select LinkFrom(W) for each Link L where L in (
               select Predecessor(P) for each NUnit P where P = U
            )
         ),
         select LinkWeight(L) for each Link V where V in (
            select Predecessor(P) for each NUnit P where P = U
         )
      ));
   end;
```

The function WeightSum has no return value and represents a procedure. The Set statement updates the activation function and uses two foreign functions sum and listprod, which represent functions giving the sum and the cross product of a list of values. The list of values are produced by select statements, which calculate the activation values of the preceding units of unit U and the weights of the respective connection links.

```
Create function WeightSumI(U NUnit) as
   begin
      Set Activation(U) = sum(listprod(
         InputData,
         select LinkWeight(L) for each Link V where V in (
            select Predecessor(P) for each NUnit P where P = U
         )
      ));
   end;
```

The function WeightSumI is similar to the original WeightSum function with the difference that the activation values are represented by the Input data stream.

```
Create function Sigmoid(U NUnit) as
   begin
      Set Activation(U)=pow(1 + exp(-Activation(U)), -1);
   end;

Create function Ident(U NUnit) as
   begin
      Set Activation(U) = Activation(U);
```

```
        end;
```

The Sigmoid and Ident function are defined for completeness. A better (faster) solution would be to provide foreign realizations. We used foreign functions for the power (pow) and the exponential (exp) operator only.

```
        Set UpdateOrder(XOR-Net) -> ((Input, 1), (Hidden, 2), (Output, 3));
        Set UpdateOrder(Input) -> ((Input1, 1), (Input2, 2));
```

By defining the UpdateOrder function the correct and controlled update order of the processing units can be guaranteed. The UpdateOrder function for the Input layer is actually not necessary and can be skipped.

3.3.2   Training Method

Now we define the backpropagation training algorithm. Due to similar reasons as with the input functions, two training methods have to be defined, one for the output layer and on for the rest of the network.

```
        Set LearnF(XOR-Net) = BackProp;
        Set LearnF(Output) = BackPropO;
```

The backpropagation algorithm consists of two passes, one for the calculation of the error value delta and one for the weight update. The error value calculation can be expressed by the following equations

$$\delta_i(l) = (t_i - a_i(L)) \cdot (a_i(L) \cdot (1 - a_i(L))), \qquad l = L$$
$$\phantom{\delta_i(l)} = (\Sigma\, \delta_k(l + 1) \cdot w_{ki}(l, L + 1)) \cdot (a_i(l) \cdot (1 - a_i(l))), \qquad l = L - 1, \ldots, 1$$

The update of the weight connections is performed according to the following formula,

$$\Delta w_{ij}(l - 1, l) = LR \cdot \delta_i(l) \cdot a_j(l - 1), \qquad LR = \text{learnrate}$$

The translation in OSQL is

```
        Create function BackProp(U NUnit) as
        begin
          Set LinkDelta(L) = y for each Link L where L in (
            select Predecessor(P) for each NUnit P where P = U
          ) and y = (sum(listprod(
            select LinkDelta(M) for each Link M where M in (
              select Predecessor(Q) for each NUnit Q where Q in (
                select LinkFrom(R) for each NUnit R where R = Q
              ) and
                Q = L
            ),
            select LinkWeight(W) for each Link W where W in(
              select Predecessor(Q) for each NUnit Q where Q in (
                select LinkFrom(R) for each NUnit R where R = Q
              ) and Q = L
            )
          ) * Activation(L)) * Activation(L) * (1-Activation(L)));

          Set LinkWeight(L) = LinkWeight(L) + x for each Link L where L in (
            select Predecessor(P) for each NUnit P where P = U
          ) and x = LearnRate (Hull U) * LinkDelta(L) * (
            select Activation(V) for each NUnit V where V = LinkFrom(L)
          );
        end;
```

The order of the evaluation of the update functions is guaranteed by the procedure, due to the dependence of the weight value on the error value.

```
Create function BackPropO(U NUnit) as
  begin
    Set LinkDelta(L) = y for each Link L where L in (
       select Predecessor(P) for each NUnit P where P = U
    ) and (
       valpos(CheckData(Hull L), Order(L)) - Activation(L)
    ) *Activation(L)*(1-Activation(L));

    Set LinkWeight(L) = LinkWeight(L) + x for each Link L where L in (
       select Predecessor(P) for each NUnit P where P = U
    ) and x = (LearnRate (Hull U) * LinkDelta(L) * (
       select Activation(V) for each NUnit V where V = LinkFrom(L)
    ));
  end;
```

The foreign function valpos(S, i) results the i-th element of data stream S. The object hierarchy of a neural unit defines a function for all units it contains. To access this function in one of the contained units, the OSQL operator Hull is introduced. X(Hull Y) delivers the next defined function X in the hierarchy above Y. This allows to define a function at a high level of the hierarchy and makes it a default value for all underlying objects. This is necessary because of the lack of recursion in OSQL queries and gives the advantage of a concise syntax.

```
Set LearnOrder(XOR-Net) -> ((Input, 3), (Hidden, 2), (Output, 1));
Set LearnOrder(Input) -> ((Input1, 1), (Input2, 2));
```

Analogous to the update order the learn order of units is defined.

After the definition of the static and dynamic components of the neural network for our XOR-example we proceed with the training phase of the network. This is described in depth in [17], so we give the OSQL calls only with brief explanations only.

We can now define the input and check data stream via select statements,

```
Set InputData(XOR-Net) = select x, y from testdata;
Set CheckData(XOR-Net) = select z from testdata;
```

apply the neural network specific Learn statement of the OSQL formalism,

```
Learn XOR-Net repeat 3000;
```

and access the result values via the output function,

```
Select OutputData(XOR-Net);
```

### 3.4 Predefined network paradigms

Many commonly known neural net paradigms are normally predefined in the database system. They are realized by more specialized object types of the NUnit type. This allows the fast and easy definition of neural network objects for a lot of known problems. To accomplish repetitive tasks, as for example the definition of processing elements, macro like procedures are predefined and made available to the user. At the end of this section we want to give the amazingly short definition of the above implemented XOR network with the specialized functions and types supported by the system.

```
Create NEUNET (name) instance XOR-Net ("Exclusive OR example");
Add type BPN to XOR-Net;
```

```
    InitializeNeuralNet(XOR-Net);

    LayerSize(XOR-Net, Input, 2);
    LayerSize(XOR-Net, Hidden, 1);
    LayerSize(XOR-Net, Output, 1);

    Set LearnRate(XOR-Net) = 4.00;
```

# 4   The Role of Neural Networks in Knowledge-based Systems

In the last few years the focus of the database research community shifted from the urge to increase the performance of systems to a beneath stronger drive to embed more intelligence into these systems. This movement took place because of the employment of "intelligent" systems to handle tasks, which were judged not suitable for computers a few years ago. The development of the performance factor of today computer systems allowed to overcome this opinion and forced a rethinking of the problem solution process for the system designer. The conventional 'static' algorithmic approach has proved viable in many applications, but lacks generally a dynamic component. That means that a static algorithm can not react sensible to situations different to its design propositions.

We, as human beings, are used to assimilate imprecise information (as fuzzy data or empirical rules) to make decisions in new (and 'unlearned') situations, based on general knowledge. Accordingly we demand that these new computational systems have to be capable to recognize, represent, manipulate, interpret and use information based on fuzzy uncertainties [1].

Key features of such systems are knowledge representation and inference mechanism. This was one of the reasons, which led to the merger of database systems and knowledge representation and called for the development of knowledge based management systems.

## 4.1   Knowledge based management systems

A knowledge based management system (KBMS) manages large and shared knowledge bases for knowledge based systems. With the realization of a KBMS two different functional levels can be distinguished [3]:

- The computational level, and
- The knowledge level

The computational level guarantees the efficient implementation of the knowledge bases and supplies efficient access and manipulation facilities to the stored knowledge representation.

The knowledge level administrates the knowledge representation scheme, which comprises development, refinement, debugging of the data model and the associated knowledge base. Further it provides tools for evaluating the adequacy, soundness, completeness.

Generally knowledge is represented by facts and rules. Facts correspond to normal data stored in conventional database systems. Rules can be seen as statements of the modelled world. Further facts can be derived by evaluation of these rules (the inference process). In other words, the intensional knowledge of rules produces further extensional knowledge as facts.

Built on top of a knowledge-based system is normally an expert system. This is a program, which exploits the inference component of the KBMS and "addresses problems normally thought to require human specialist for their solution" [5].

An expert system can be better defined by its functionality, which comprises 3 important activities ("the 3 I's"): inference, interaction, information.

*Inference*: Inference allows to draw conclusions out of the knowledge base without knowing all possible information.

*Interaction*: The program can interactively ask for more information to find a (better) solution or to avoid uncertainty.

*Information*: The program has to justify its inference and gives information to the user, how it reached the conclusion.

Expert system proved extremely well in the last few years and can be seen as the biggest accomplishment of artificial intelligence.

## 4.2   Advantages of database technology

The classic expert systems were stand-alone implementations with their own knowledge base administration modules. It was therefore a straightforward development to embody database systems to accomplish the task of data (in our case knowledge) administration. The transition of a database system to a knowledge based management system results in a number of important advantages, as persistency, interactive query interface, integrity control, concurrency management, recovery, and secondary data management [8].

Generally the given features comprise the capability of a database system to manage shared data independently of application programs [12].

## 4.3   Neural network knowledge representation

In the area of traditional knowledge engineering the hypothesis is inherent that knowledge is represented symbolically by expressions and data structure. It is called the knowledge representation hypothesis, which means that the evaluation and the manipulation of this knowledge produces subsequently new results (further knowledge).

Neural networks allow a different approach to the meaning of knowledge. The representation of knowledge can be seen as the processing of the neural network [13]. The global rules are not required, but are inherently part of the processing in the individual processing nodes (similar to the behaviour of neurons).

We do not argue that it is wise to substitute the conventional syntactical rules by neural network processing systems, but we claim that neural networks have to be included into the framework of knowledge representation.

Quite a number of paper appeared in the last few years, which used neural network as representations of intensional knowledge. A typical example is the evolving area of information retrieval, where neural networks proved very valuable [2]. In practice neural networks are useful in a large number of applications. They have qualified themselves as an instrument for handling problems, which are to complex or to costly for a conventional algorithmic solutions.

## 4.4   Neural networks and expert systems

The intensional knowledge inherent in neural networks makes them extremely useful for the use in expert systems. Basically 2 reasons for the employment can be distinguished [7],

- Automatic construction of the knowledge bases, and
- Elements in the knowledge bases

### 4.4.1 Automatic knowledge base construction

Conventional expert systems suffer from the problem of the very expensive and time consuming task to create the knowledge base. It is difficult and error prune even for a human expert to express his information in "if-then" rules and to attribute confidence values. The rule set is therefore often incomplete, inconsistent and incorrect. It needs a long time after the creation to adapt and correct the knowledge base under the guidance and surveillance of a number of experts.

Much more efficient would be to provide data (large amounts of it, if necessary) to the known training algorithms of neural networks. These algorithms can create automatically the rules for the expert system. This is extremely useful, if the amount of data is very large and noisy.

### 4.4.2 Knowledge base elements

Another possibility is the use of neural networks as elements of the knowledge base. This can be in addition to conventional rules or alone as the basic rule type.

Two different types of neural network elements can be distinguished in a knowledge base, "if-then" nets (rules) and pattern matching nets.

*"if-then" rules*: Neural networks can be used to express the commonly known "if-then" rules as well. These rules solve one inherent problem of neural networks in expert systems, the explanation of the inference process. Two criteria have to be satisfied by these rules,

- *Validity*: for any possible instantiation of variable (also variables, which are not input to the rule) the rule must hold.
- *Generality*: if any condition of the rule is deleted, the rule must fail (in the literature this is also called maximal generality).

Further "if-then" rules are very helpful with the construction of expert systems under the guidance of human experts to express knowledge.

*Pattern matching nets*: A number of problems are hard to solve by "if-then" rules, but easily by neural networks. These are problems in the area of pattern recognition or statistics. These nets have the advantage that they are relatively resistant against noise, which makes them extremely suitable for a large number of practical applications.

However the training algorithm finds an association between the input and output values only, without an explanation for the correlation. This results in a poor explanation (introspection) capability, which is insufficient for expert systems.

## 4.5 Integration of neural networks and KBMS

Many existing neural network systems employ neural network techniques by using a connected but separate neural network simulator. This is based on the fact that no knowledge based system exist, which supports and embeds neural networks directly in the general knowledge representation framework. This situation leads to a number of problems.

The main problem is a lack of expressional and formally sound mechanisms for the user within his application. In many cases two (mostly) different designed systems exist (the application and the neural network system) and an artificial and often clumsy interface. The user is often heavily restricted two a few general network paradigms or dependent on small hooks in the application for the neural network employment. This situation led in the past to the development of non-standardized, proprietary and specialized systems, which were only applicable to a very restricted range of problems.

Further problems to mention are often the complexity, the inflexibility and the lacking performance of the systems.

The obvious solution to this problem is the embedment of neural networks directly into the generalized knowledge environment of a KBMS employing the proposed integration framework.

In Figure 2 the generic structure of an integration approach is given:

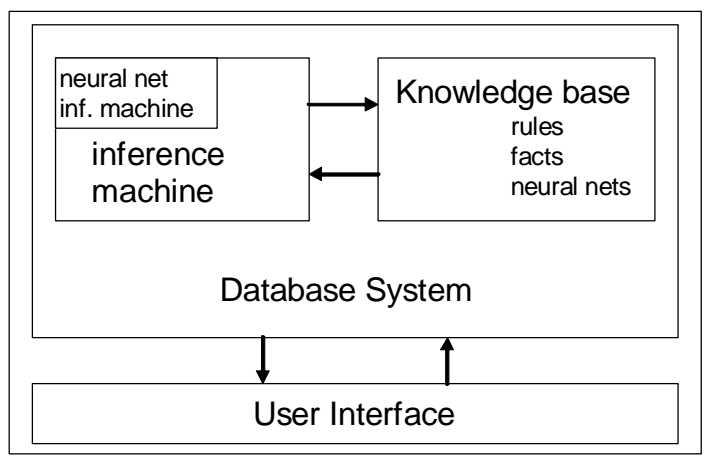

Figure 2: Integration framework

To be successful, the embedment of neural networks into the framework of a KBMS has to be performed on both internal levels of the KBMS, the knowledge and computational level.

4.5.1  Computational level

The computational level has to provide a generic neural network type. This type can be used to instantiate neural network objects, according to user definable network paradigms. The administration of these objects (like access, update, deletion etc.) is naturally and homogenously supported by the database facilities of the KBMS.

For the handling of alternate states of a neural network the versioning mechanism of the KBMS can be used. This allows easily to handle different versions of neural networks of one paradigm for different problems or networks of different paradigms for one problem.

A necessary property is the capability to update or change the characteristics of a network. It is not comprehensive enough to change weights or activation values, but it must be possible to change the topology of a network. This situation arises with the creation of the knowledge base for a neural network based expert system. The connectionist training algorithm produce new networks or change existing ones according to training data. The latter is called the expandable network problem [7] in the literature. New intermediate cells are created in a network to fit a problem better. Generally, whole rule sets can be generated from a neural network, which approximate a collection of "if-then" rules.

This leads to the necessity of extendibility of a network type in a KBMS.

4.5.2   Knowledge level

The knowledge level provides the behavioural component for the neural network, like the update and the training method of the network. Further it embeds the neural network into the knowledge base framework of the KBMS.

Trained neural networks are similar to rules in the conventional symbolic sense. This allows using and handling them on the knowledge level with the conventional methods. The training of a neural network has not to be seen different from the knowledge contained in the symbolic rules, according to the classification. Therefore the training phase of the neural network is intensional knowledge and fits smoothly into the knowledge representation framework.

A further important phase is the debugging of an existing expert system. Debugging is always necessary if a new rule is inserted into or an existing rule is changed in the knowledge base. This can generate an error in the rule system, which is not apparent because of the complex interrelationship of rules and confidence factors. With an neural network expert system the debugging can be automated. When training examples are added or modified, the training algorithm itself keeps the knowledge base consistent. The algorithms can detect easily the rules, which contradict the updated training examples.

4.5.3   Interaction of knowledge and computational level

The knowledge and computational level are neither totally separate nor stand-alone. A strong interaction is necessary to accomplish the different knowledge representation tasks.

To train a neural network a large number of training examples is necessary. These examples have to be available in the knowledge level and must be provided by the computational level too. In this situation the database system shows its power. It administrates easily large data sets and provides an efficient access to them. Further the interface languages of modern database systems allow to establish logical views to the stored information. This makes it easy to analyze training data sets and to restrict them to the interesting components. This also gives the possibility to aggregate or arrange the information, which in many cases can ease and speed-up the training phase of a neural network significantly.

Not only the data sets alone are administrated by the computational level, but also the neural networks as basic objects of the database system. This allows the usage of efficient access mechanisms of the database system to handle the (in many situation) very large number of different neural network objects. This results into two advantages, a fast access to a neural network and a high level logical specification of similar network sets.

# 5   Conclusion and Acknowledgement

We presented an integration framework for neural networks into database systems using the object-oriented approach. We gave a justification by a practical realization using the Iris database systems. Finally based on these findings we laid out an architecture for the embedment of neural networks as knowledge component into knowledge-based management systems.

This paper[1] comprises work done by the author over the last ten years. The ideas presented were deeply influenced and motivated by the intellectual discussion with my academic "father" Professor Günther Vinek whom I want to express thoroughly my gratitude for his guidance and support.

Based on the described framework several running systems were developed by the author, as NeuDB [18], NeuroAccess [19], NeuroWeb [20], and N2Grid [21].

---

[1] This paper is an extended version of [16] which was awarded with the "Outstanding Paper Award" of the Society for Computer Simulation.